\documentclass[prx,twocolumn,english,longbibliography,superscriptaddress]{revtex4-2}
\usepackage{ifthen}
\usepackage{color}
\usepackage{graphics,graphicx,epsfig}
\usepackage{epsf,epstopdf,wrapfig}
\usepackage{amssymb,amsfonts,amsmath}
\usepackage[export]{adjustbox}
\usepackage[english]{babel}
\include{epsf}
\usepackage{ifthen}		

\newcommand{\gen}{{\text {gen}}}
\newcommand{\post}{{\text {post}}}
\newcommand{\Treg}{{\text {Treg}}}
\newcommand{\Tconv}{{\text {Tconv}}}
\newcommand{\beqn}{\begin{eqnarray}}
\newcommand{\eeqn}{\end{eqnarray}}
\newcommand{\beq}{\begin{equation}}
\newcommand{\eeq}{\end{equation}}

\newcommand{\Q}{{\mathcal{Q}}}

\definecolor{junglegreen}{rgb}{0.16, 0.67, 0.53}
\definecolor{myrtle}{rgb}{0.13, 0.26, 0.12}
\definecolor{lincolngreen}{rgb}{0.11, 0.35, 0.02}
\definecolor{forestgreen}{rgb}{0.13, 0.55, 0.13}

\newcommand{\G}{\color{black}}
\newcommand{\GI}{\color{black}}

\usepackage[normalem]{ulem}

\begin{document}

\onecolumngrid
\begin{center}
{\large \bf Variability in the local and global composition of human T-cell receptor repertoires during thymic development across cell types and individuals}\\
\vspace{0.5cm}

Giulio Isacchini$^{1,2,*}$,  Valentin Quiniou$^{3,4}$, H\'el\`ene Vantomme$^{3,4}$, Paul Stys$^3$,  Encarnita Mariotti-Ferandiz$^{3,\dagger}$, David \\ Klatzmann$^{3,4,\dagger}$, Aleksandra M. Walczak$^{2,\dagger,\ddagger}$, Thierry Mora$^{2,\dagger,\ddagger}$, and Armita Nourmohammad$^{1,5,6,7,8,\dagger,\ddagger}$\\
\vspace{0.3cm}
{\it $^1$Max Planck Institute for Dynamics and Self-organization, Am Fa\ss berg 17, 37077 G\"ottingen, Germany\\
$^2$Laboratoire de physique de l'\'ecole normale sup\'erieure (PSL University), CNRS, Sorbonne Universit\'e,\\ and Universit\'e de Paris, 75005 Paris, France\\
$^3$Sorbonne Universit\'e, INSERM, Immunology-Immunopathology-Immunotherapy (i3),\\ F-75005, Paris, France\\
$^4$AP-HP, H\^opital Piti\'e-Salp\^etri\'ere, Biotherapy (CIC-BTi), F-75651, Paris, France\\
$^5$Department of Physics, University of Washington, 3910 15th Avenue Northeast, Seattle,\\ WA 98195, USA\\
$^6$Paul G. Allen School of Computer Science and Engineering, University of Washington,\\ 85 E Stevens Way NE, Seattle, WA 98195, USA\\
$^7$Department of Applied Mathematics, University of Washington, 4182 W Stevens Way NE,\\ Seattle, WA 98105, USA\\
$^8$Fred Hutchinson Cancer Center, 1241 Eastlake Ave E, Seattle, WA 98102, USA\\
}
\end{center}
\vspace{0.2 cm}
\leftskip=0.5in 
\rightskip=0.5in 

{\small
\noindent$*~$Current affiliation: Peter Debye Institute for Soft Matter Physics, Leipzig University, Leipzig, Germany\\
$\dagger~$Co-senior author\\
$\ddagger~$These authors contributed equally. Correspondence should be addressed to: aleksandra.walczak@phys.ens.fr, thierry.mora@phys.ens.fr, and armita@uw.edu.\\}
\vspace{0.5cm}

{The adaptive immune response relies on T cells that combine phenotypic specialization with diversity of T cell receptors (TCRs)} to recognize a wide range of pathogen{\GI s. TCRs are acquired and selected during T cell maturation in the thymus. Characterizing TCR }repertoires across individuals and {\GI T cell } maturation stages is important for {\GI better understanding adaptive immune responses and for developing } new diagnostics and therapies. {\GI Analyzing a dataset of human TCR repertoires from thymocyte subsets, we } find that the variability between individuals generated during {\GI the TCR }V(D)J recombination is maintained through all stages of T cell maturation and differentiation. The {\GI inter-individual variability } of repertoires of the same cell type is of comparable magnitude to the variability across cell types within the same individual. To zoom in on smaller scales than whole repertoires,   {\GI we defined a distance measuring} the relative overlap of locally similar sequences in repertoires. We find that the whole repertoire models correctly predict local similarity networks, suggesting a lack of forbidden T cell receptor sequences. The local measure correlates well with distances calculated using whole repertoire traits and carries information about cell types. \\

\vspace{0.5cm}

\leftskip=0.in 
\rightskip=0.in 
\twocolumngrid

\section{Introduction} 
{\GI The T cell adaptive immune response leverages various cell subsets. CD8+ cells take on mostly a cytotoxic role,  i.e.  killing infected cells. CD4+ cells differentiate into two subsets, regulatory (Treg) and conventional (Tconv) cells. Tconvs acquire effector helper function and help coordinate the immune response upon activation in the periphery. Tregs modulate the immune response by down-regulating the activity and response of different cells in the immune system. 

To perform their distinct functions, CD8+ and CD4+ cells react to different families of the MHC molecule (class I and II, respectively). Therefore, the selective pressures exerted on their associated TCR repertoires are believed to be markedly different. Indeed, previous studies \cite{EMERSON201314, Li2016a, Carter2019,Isacchini2021a} have described differential properties of the CD4 and CD8 repertoires, reporting significant yet limited statistical differences, mostly related to the V and J gene usage. Similarly, small statistical differences have been reported in the repertoires of conventional (Tconv) versus regulatory (Treg) CD4+ T cells \cite{Isacchini2021a,Lagattuta2022}. However, despite a recent study comparing the repertoires of different thymic subsets in mice \cite{Camaglia2022}, it is still unclear how thymic selection shapes the repertoires of distinct T cell subsets, and how these selection forces vary across individuals.}

To recognize the large variety of possible antigens, T cells express a broad diversity of T-cell receptors (TCR) generated by the random rearrangement of their $\alpha$ and $\beta$ chains. Because this process is stochastic, it may generate receptors with undesirable properties, which must be vetted \cite{Yates2014}.
During their development in the thymus, T cells undergo a selection process that promotes receptors with good affinity to the Major Histocompatibility Complex (MHC), to make sure that they have the minimum necessary recognition properties (\textit{positive selection}). At the same time, receptors with {a} too strong affinity to self-peptides presented by the MHC are less likely to be released into the periphery,  to limit an immune response against the self (\textit{negative selection}).
Since the MHC is highly polymorphic \cite{murphy2008janeway}, this process is expected {\GI to be at least in part specific to }each individual.
Understanding how these processes shape the TCR repertoire and characterizing the heterogeneity within and between individuals, is key to {\GI a} better understanding of the fundamental aspects underlying auto-immunity as well as immune disorders.

High-throughput sequencing of TCR repertoires \cite{Hou2016,Georgiou2014,Bolotin2015a,Mcdaniel2016,DeKosky2013a, Turchaninova2013c,Dekosky2014} has made it possible to study differences between different individuals as well as T cell subsets, giving insight into their respective selection processes. T cells differentiate in the thymus into two major subpopulations of cells depending on whether they express the CD4 or CD8 receptor on their surface. 

{Functionally, CD8+ and CD4+ cells take on different roles. CD8+ cells take on a cytotoxic role and they kill infected cells. CD4+ cells differentiate into two subsets, regulatory (Treg) and conventional (Tconv) cells. Tregs modulate the immune response by down-regulating the activity and response of different cells in the immune system. Tconvs acquire effector helper function and help coordinate the immune response upon activation in the periphery.} 

Because TCR repertoires are so diverse, typical samples still show substantial variability, 
 even when they are from the same subset and same individual~\cite{BensoudaKoraichiPNAS2023}. Comparisons between repertoires must be done at an aggregate level, by identifying statistical features that discriminate between them. A general strategy for this task is to define generative models of TCR sequences for each subset, and compare them using measures of divergence between their distributions \cite{Sethna2020,Isacchini2021a}. We refer to this approach as {\em global} repertoire comparison.
A previous study  \cite{Sethna2020} of bulk TCR repertoires from blood samples showed only moderate variability across individuals, largely driven by differences in the statistics of the recombination process~\cite{Tonegawa1983,Davis1988} rather than individual-specific selection, {which contrasts with the idea that HLA drives the form of the repertoires}.
It remains unclear to what extent  the population variability in the recombination process impacts 
our ability to detect  the  differences between  T cell subsets due to functional selection.

Model-based comparisons, even when they rely on powerful neural networks, are usually dominated by some key statistics of the data, such as germline gene usage, or CDR3 length and position-dependent amino-acid usage. A possible concern is that they can miss  important differences in repertoire properties, such as the depletion or enrichment of particular sequence motifs, which would strongly affect the structure of the repertoires locally in sequence space but may only have a limited impact on their global features as captured by the generative models. 
Previous studies have used local network measures of similarity between closeby sequences to detect immune stimuli or phenotypes shared by TCRs \cite{Pogorelyy2018b,TCRNET,ALICETCRNET,Dash2017,Blackwell2021,Glanville2017,Huang2020}, suggesting that local differences may be important for discriminating between repertoires beyond global statistical features. We refer to such approaches as {\em local} repertoire comparisons.

We applied both global and local comparison methods to TCR repertoire data obtained from the thymi of 9 organ donors, sorted into functional T cell subsets at different levels of thymic maturation. By comparing these different subsets using a combination of model-based and local network analyses, we study how the T cell repertoire evolves across the various stages of thymic development and quantify the heterogeneity between individuals.

\begin{figure*}[ht!]
\begin{center}
\includegraphics[width=1.\linewidth]{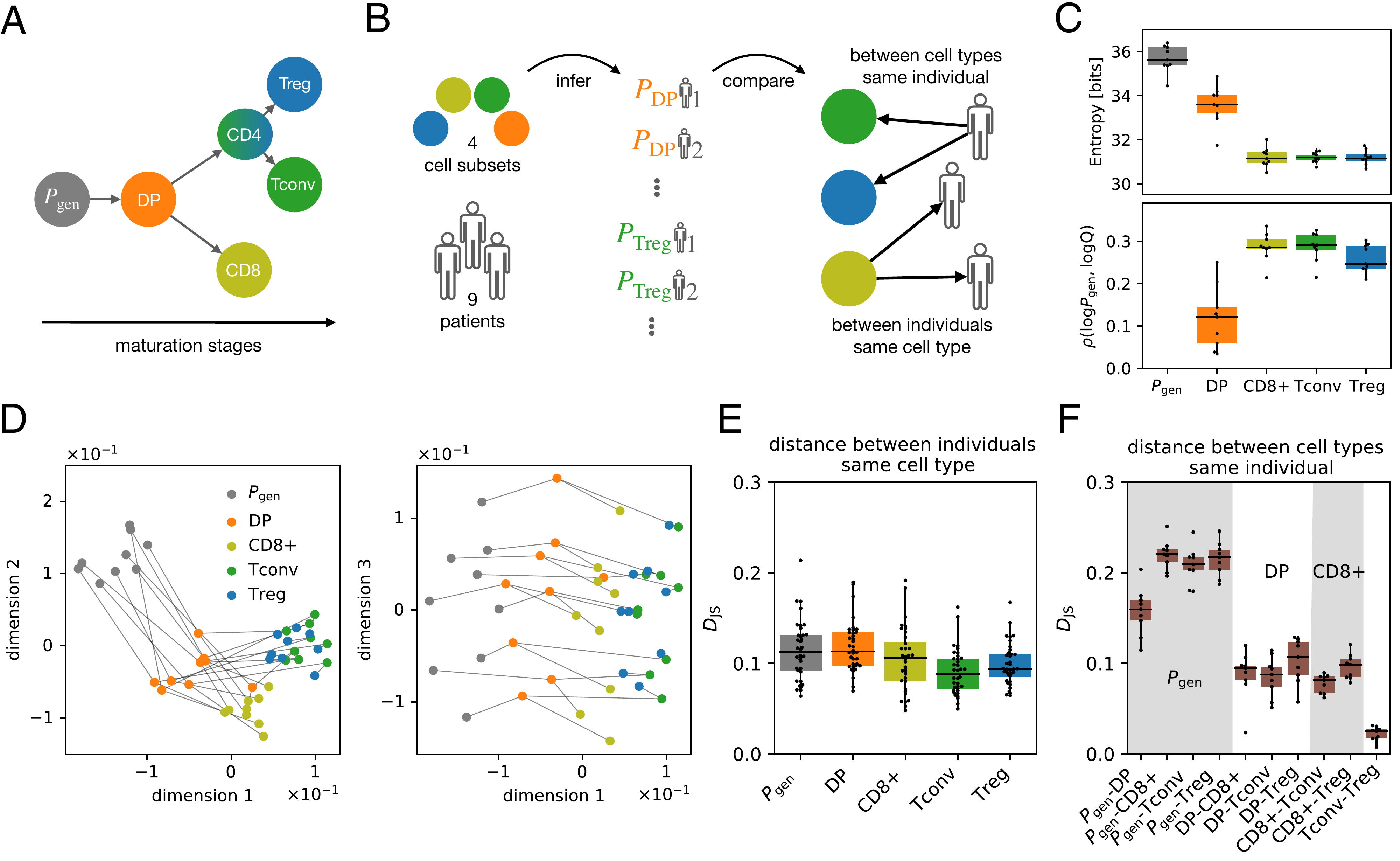}
\caption{
{\bf  Comparison of global repertoire features across individuals and thymic development stages.} {\bf(A)} Schematic of the thymic development stages at which repertoires were sequenced. $P_{\rm gen}$ denotes the output of the raw VDJ recombination process. Sequences are initially selected into the double positive (DP) pool, and further differentiate into single positive phenotypes: CD8+, and conventional and regulatory CD4+.
  {\bf(B)} Analysis workflow. A model is inferred for each individual and cell subset, in addition to the $P_{\rm gen}$ model inferred from unproductive sequences. Models are then compared across cell types and individuals.
  {\bf(C)} Entropy (top) and Pearson correlation between the logarithms of the generation probability and selection factor (bottom), as a function of {\GI the }maturation stage.
  {\bf(D)} 
  Low dimensional projection of distances between repertoires. Dimensionality reduction was performed using Multi-dimensional Scaling of the distance matrix defined by the Jensen-Shannon divergence $D_{JS}$ between the inferred model distributions.
  {\bf(E)}
$D_{\rm JS}$ between repertoires of the same cell type but from different individuals.  {\bf(F)} $D_{\rm JS}$ between repertoires from the same individual but of different cell types.}
\label{fig:klatz1}
\end{center}
\end{figure*}

\section{Results}

\subsection{Generative models of individual T-cell subsets}

We analyse high-throughput TCR$\beta$ repertoire sequence data of purified T cell subsets from different stages of maturation (see Fig.~\ref{fig:klatz1}A for a schematic of the different stages). We analysed purified samples of CD4+CD8+CD3+ (double positive: DP), CD3+CD8+ (CD8+),  CD3+CD4+CD25+ (Treg), CD3+CD4+CD25- (Tconv) from thymic samples from $n=9$ individuals. Details on sorting strategies and high-throughput sequencing steps are described in \cite{Klatzmann2020} where part of this dataset was presented. DP cells represent an early stage of development after successful recombination and {\G selection of a receptor, prior to commitment to any functional fate.} The statistics of their repertoire should closely follow that of the recombination process. During selection, cells with higher affinity to one of the two major classes of MHC differentiate into either CD8+ or CD4+ cells {\G (Treg and Tconv)}. 

We used the IGoR software \cite{Marcou2018} to infer a VDJ recombination model (called $P_{\rm gen}$) for each of the n=9 individuals from the unproductive TCR sequences pooled from all 4 subsets. Unproductive sequences are fossil records of unsuccessful recombination events on the second chromosome and {\GI are }thus believed to be free of selection effects that impact productive sequences. To accurately account for selection, we then trained a SONIA model \cite{Sethna2020} on productive sequences for each subset, $P_{\rm DP}, P_{\rm CD8}, P_{\Tconv}, P_{\Treg}$, and each individual ($9\times4$ models in total). This model is built on top of the VDJ recombination model, $P=Q\times P_{\rm gen}$, where $Q$ is a subset-specific selection factor dependent on receptor sequence features, including V-, J- gene usages, and junction length and amino acid composition.
As previously done in \cite{Isacchini2021a}, we can {characterise the diversity of repertoires by evaluating the entropy of the inferred models and estimate the similarities between the repertoires by computing the Jensen-Shannon divergence of the models} (see Fig.~\ref{fig:klatz1}B for a summary of the analysis strategy).

\subsection{Thymic selection reduces sequence diversity by amplifying biases of VDJ recombination}

We first quantified how the effective diversity of the repertoire changed throughout thymic development, by computing the Shannon entropy of each inferred model (Fig.~\ref{fig:klatz1}C, top). This entropy {\G is a measure of {\GI the }diversity of TCRs at different developmental stages and can be thought of as the number of distinct TCR sequences. It  is different from} 
 the raw number or clonality of TCR, both of which are subject to sampling biases. 
Following VDJ recombination, when receptors are well described by the $P_{\gen}$ distribution, the entropy is the highest at around 36 bits, 
 corresponding to an effective sequence space of size $2^{36}\approx 10^{11}$. Entropy is significantly reduced at the DP stage (34 bits $\simeq 10^{10}$ sequences), and even further at the single positive stages of  CD8+, Tconv, and Treg (31 bits $\simeq 10^9$ sequences).

We asked whether this reduction of diversity  results from existing biases in the VDJ recombination process, or is independent of them. The bottom part of Fig.~\ref{fig:klatz1}C shows that entropy reduction is accompanied by an increasing correlation between $P_{\rm gen}$ and $Q$ as maturation progresses, meaning that selection reinforces heterogeneities already present at generation, and thus reduces diversity through the ``rich get richer'' effect. This observation is consistent with reports of a similar correlation in fully matured TCR sampled from blood~\cite{Elhanati2014}. The results in Fig.~\ref{fig:klatz1}C  show the progression of this correlation during thymic selection, with onset as early as the DP stage.

\subsection{Global comparison of repertoires between individuals and thymic maturation stages}

We then estimated the similarity between repertoires at various stages of thymic development and in different individuals, by computing the Jensen-Shannon divergence $D_{JS}$ (an information-theoretic measure of distance between distributions) between the inferred models. The resulting distance matrix
(Fig.~S1A) displays a complex structure, which is better interpreted by projecting repertoires into a distance-preserving low-dimensional map using multidimensional scaling (MDS) \cite{MDS}.
We find that 3 embedding dimensions are sufficient to describe the main properties of the distance matrix (Fig.~S1B). {To effectively visualize these embeddings in 2 dimensions,}  we rotated these 3 main axes  in Fig.~\ref{fig:klatz1}D such that the third dimension best aligns with the identity of individuals, which is tracked across subsets by gray lines. We observe a progression of thymic differentiation along dimensions 1 and 2, with clusters corresponding to each stage, while dimension 3 delineates individuals. {\G Notably, the VJ gene features alone cannot separate the intra-individual cell subsets (Fig.~S6B).}
This result suggests that individual heterogeneity imprinted by initial differences in VDJ recombination remains frozen throughout the maturation process,  in agreement with previous analyses on unsorted repertoires \cite{Sethna2020}.

An alternative way to represent similarities between repertoires is to find a common {\G encoding} space for \textit{sequences} (by using neural networks, as discussed in the Methods) instead of whole repertoires, and then  directly compare the location of different receptor sequences from different datasets (Fig.~S2). While sequence variability is large and no clear separation of subsets is visible at the individual sequence level,  we find the average location of receptor sequences {\G in the encoding space } to follow  the expectations from the differentiation process ({\G Fig.~S2), consistent with the global repertoire analy{\GI s}is in Fig.~\ref{fig:klatz1}D}.

{\G In Figs.~\ref{fig:klatz1}E,~\ref{fig:klatz1}F we observe that }  variability of the same cell type between individuals is quantitatively comparable to that between cell types within a single individual. A more detailed quantification of differences across individuals with fixed cell types (Fig.~\ref{fig:klatz1}E) shows that individual differences are stable across stages of maturation. Differences across cell types (with fixed individuals), shown in Fig.~\ref{fig:klatz1}F, follow the known hierarchy of thymic development, with the recombination model $P_{\rm gen}$ furthest from all subsets, but closer to DP, and single positive stages being equidistant from DP and from each other, with the exception of conventional and regulatory CD4+ cells, which are very similar.
\begin{figure*}[ht!]
\begin{center}
\includegraphics[width=.85\linewidth]{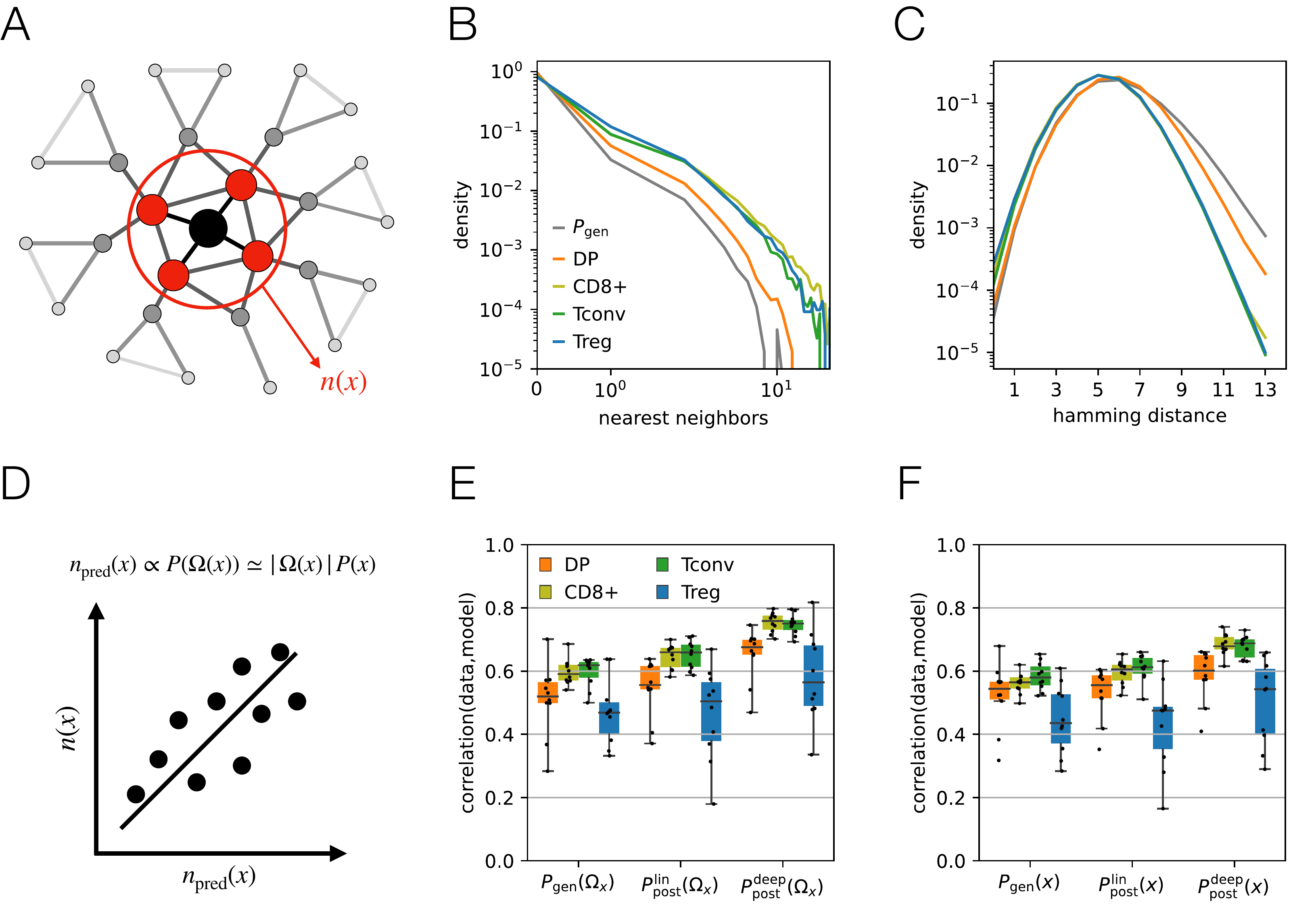}
\caption{{\bf Local properties of a repertoire} {\bf(A)} Nearest neighbours of a sequence $x$ are defined as sequences that differ by only one aminoacid substitution. {\bf(B)} 
The probability that a sequence has exactly $n$ nearest neighbours with the same $VJ$ genes is shown for different repertoire subsets.  As T cells experience selection the average number of nearest neighbours increases. The distribution is well reproduced by a synthetic repertoire sampled from the model $P_{\rm post}$; see Fig.~S3.
{\bf(C)} Probability that two randomly sampled sequences with the same $VJ$ gene combination have Hamming distance $d_H$ between their aminoacid sequences is shown for different repertoire subsets.  During thymic development, the average sequence distance decreases. {\GI The c}olor code is similar to (B). {\bf(D)} We quantify the accuracy of our model by computing the pearson correlation between {\GI the }predicted and {\GI the }observed number of neighbours {\GI over the 10 most probable VJ combinations}. {\bf(E)} The {Pearson correlation} between the true value for the  number of nearest neighbours and the estimated values based on the model  $n_{\rm pred}(x)$, for $P_{\rm gen}$,$P^{\rm lin}_{\rm post}$ and $P^{\rm deep}_{\rm post}$ models.  {\bf(F)} Similar to (E) but for an approximate predicted number of neighbors  $\tilde n_{\rm pred}(x)$, for which the   model probabilities are evaluated using only the center sequence, instead of all the sequences in a local neighbourhood. This approximation relies on the smoothness of the probability landscape.  The prediction of the estimators remains significantly correlated with the observed number of neighbours in the data.
}
\label{fig:klatz2}
\end{center}
\end{figure*}
\subsection{Repertoires become concentrated as they mature in the thymus}

While the previous analysis gives a general bird eye's view of differences between repertoires, it is not informative about differences at the local sequence level.
Following~\cite{Klatzmann2020} we asked whether the sequence neighborhood of TCR sequences carried signatures of thymic selection. We define as ``neighbors'' sequences that have the same V and J genes, junction length, and differ by at most one amino acid {\G (Hamming distance of 1)}; see Fig.~\ref{fig:klatz2}A. The larger the overall number of neighbors, the more ``concentrated'' is the repertoire. 
Fig.~\ref{fig:klatz2}B shows the distribution of the number of neighbors of an arbitrary TCR from the repertoire, at different stages of T cell development.
\begin{figure*}[ht!]
\begin{center}
\includegraphics[width=.85\linewidth]{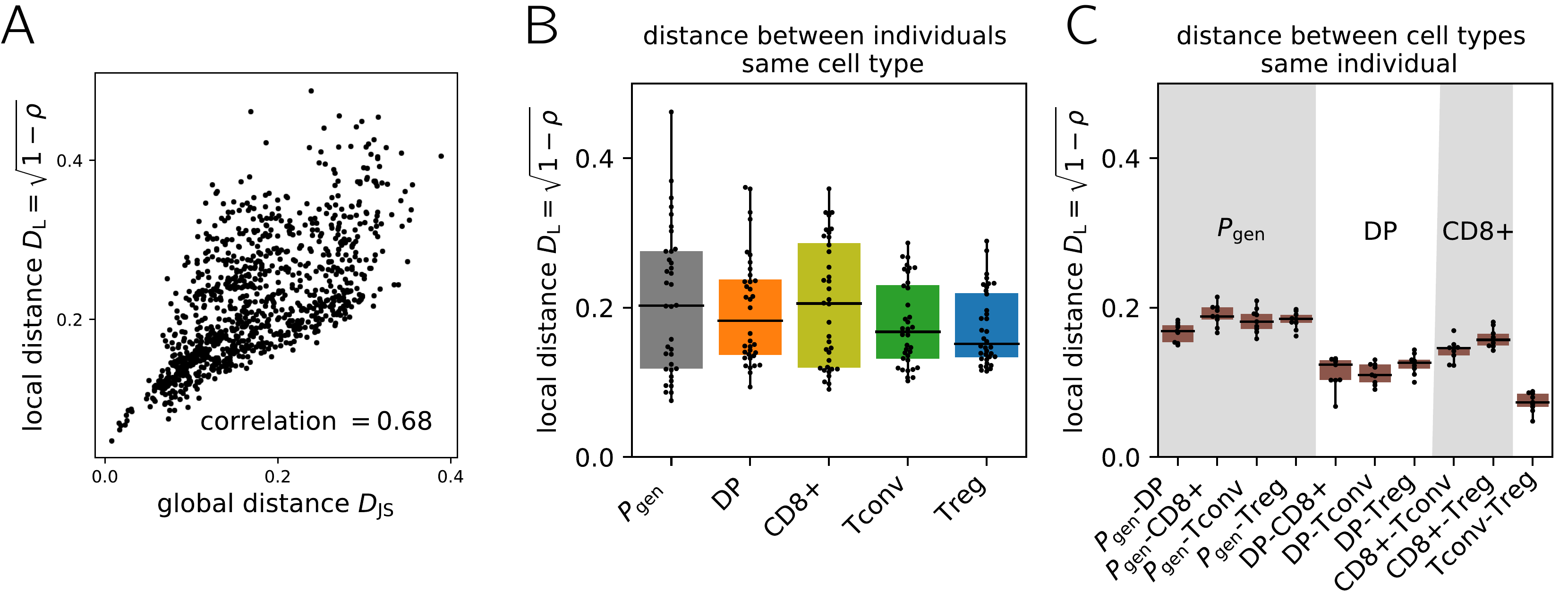}
\caption{
{\bf Comparison of local repertoire features across individuals and thymic development stages.} {\bf(A)} Local distance $D_{\rm L}$ and global distance $D_{JS}$ (Fig.~\ref{fig:klatz1}) are significantly correlated.
{\bf(B)} Local Distance $D_{\rm L}$ between repertoires of the same cell type but from different individuals.
  {\bf(C)} Local Distance $D_{\rm L}$ between repertoires from the same individual but of different cell types.
} 
\label{fig:klatz3}
\end{center}
\end{figure*}
The mean number of neighbors increases during thymic development, indicating that the repertoire gets increasingly concentrated around some preferred sequences.  In other words, selection during thymic development amplifies neighboring sequences within certain regions of the sequence space, while depleting others, which results in repertoires with reduced overall diversity but with  local 
agglomeration of sequences.

An alternative way to measure repertoire concentration is to look at the distribution of distances between any two receptor sequences (Fig.~\ref{fig:klatz2}C). Here, we consider the Hamming distance (number of unmatched amino acids in the junction) between sequences with the same V and J gene and junction length.
Receptors tend to have on average lower Hamming distances following selection, for both CD4+ and CD8+ cells (Fig.~\ref{fig:klatz2}B).  Together, these observations are consistent with the results of \cite{Klatzmann2020} on  the CD8+ T cells obtained from the same datasets, and extend them to CD4+ T cells.

Although we observed that the diversity in repertoires decays with maturation (Fig.~\ref{fig:klatz1}C), this global pruning of sequences did not necessarily imply that the pairwise distances between the receptors in a subset should decay with maturation.
We asked whether this local behaviour could be reproduced by our generative models. We pooled data from all patients for each subset to collect enough sequences to be able to train a soNNia (neural-network based) model~\cite{Isacchini2021a} for the selection factor $Q$ for each maturation stage (see Methods).
We find that both the distributions of {\GI the }number of neighbors and {\GI the }Hamming distances are well reproduced by the models (Fig.~S3), indicating that these local neighborhood differences are well captured despite the models being trained on global features of the repertoires.

We then asked whether the model could predict the number of neighbors $n(x)$ of a particular sequence $x$, and not just their distribution. Assuming that each sequence is drawn at random independently of others, the number of neighbors should be distributed according to a Poisson distribution of mean \cite{Pogorelyy2018b,TCRNET}:
\begin{equation}
n_{\rm pred}(x)=\frac{N_{VJ}}{P(V,J)} \sum_{y \in \Omega(x)} P(y) \simeq \frac{N_{VJ} |\Omega(x)| }{P(V,J)} P(x)
\label{eq:rate}
\end{equation}
where $P(x)$ is the model distribution over sequences $x$, $P(V,J)$ is the probability of picking a particular VJ pair estimated from the model, $N_{VJ}$ is the observed number of sequences with that VJ pair, and $\Omega(x)$ is the set of all potential (not necessarily extant) neighbors of $x$.
The approximation in the second equality relies on the assumption that the probability landscape is smooth, so that the probability of a neighboring sequence is on average the same as the focal sequence.

The accuracy of models can be evaluated by calculating the correlation between $n(x)$ and $n_{\rm pred}(x)$ across all sequences with $>3$ neighbours (Fig.~\ref{fig:klatz2}D). {\GI Fig.~\ref{fig:klatz2}E compares the performance of three models across cell subsets: the naked recombination model $P_{\rm gen}$, the linear (SONIA) and the  neural-network (soNNia) selection models. The performance of the soNNia model reaches the Spearman correlation of $\rho\sim 0.75$ for CD8+ and Tconv subsets. The Treg and DP datasets have smaller sizes and contain fewer nearest neighbours on average (Fig.~S7),  resulting in a nosier comparison to the model, which reduces the absolute correlation. The smooth landscape approximation only moderately degrades predictability (Fig.~\ref{fig:klatz2}F), while showing much faster computation times (by a factor  of 19 times the mean junction length), thanks to the fact that it does not require computing the probability of each neighboring sequence. These observations carry over to the prediction of the number of second neighbors (i.e., sequences with at most 2 amino acid differences) and beyond (Fig.~S4).}

Overall, these results show that the local properties of individual repertoires are well captured by the model and that the probability landscape of finding receptors sequences is relatively smooth as a function of sequence distance.  This implies that we cannot identify forbidden regions in the space of receptor sequences. 

\subsection{Local differences between repertoires}
Next, we asked how the neighborhood structures differed between cell types and individuals at the level of single sequences. For a given sequence $x$, we want to measure differences in the local network structure of nearest neighbors across different repertoires. We quantify these differences by defining a distance  based on the spearman correlation between the number of neighbors of each sequence in the two datasets:
\begin{equation}
D_{\rm L}(D_1,D_2)=\sqrt{1-\rho(n_{D_1}(x),n_{D_2}(x))}
\label{eq:local_distance}
\end{equation}
where $n_{D_1}(x)$ and $n_{D_2}(x)$ are the number of neighbors of sequence $x$ in repertoires $D_1$ and $D_2$.

Compared to the Jensen-Shannon Divergence $D_{JS}$, this distance is based on the local sequence information only, and in principle is model-independent. However, its accuracy strongly depends on sequencing depth, because it relies on counting the number of neighbors for each sequence,  see Fig.~S5A. To overcome that difficulty,  we used  the estimators $n_{\rm pred}$ evaluated with the inferred models of each repertoire to compute the $D_L(D_1,D_2)$.

When applied to all possible pairs of repertoires, the resulting local distance $D_{\rm L}$ correlates well with the global distance between repertoires $D_{JS}$ (Spearman $\rho=0.68$); see Fig.~\ref{fig:klatz3}A. Figs.~\ref{fig:klatz3}B,~C show this local distance between individuals and between cell types. These statistics of local distances closely mirror those of global distance shown in Figs.~\ref{fig:klatz1}E,~F. This means that differences in the local sequence structure of  the repertoire follows global differences captured by the models, and that the global models fully captures the local statistics.

\section{Discussion}

Previous work \cite{EMERSON201314,Li2016a,Carter2019,Isacchini2021a} has characterized global statistical differences between the TRB repertoires of fully mature CD4+ and CD8+ T cells. Here we tracked how these differences emerge during thymic development. We find that the diversity of the repertoire shrinks as a result of thymic selection. This reduction occurs through the concentration of the repertoire around particular regions of the sequence space, consistent with the findings of \cite{Klatzmann2020} on CD8+ repertoires from the same dataset. These favored sequences are typically likely to be generated by VDJ recombination (high $P_{\rm gen}$) even before thymic selection, suggesting that the recombination process has evolved to produce sequences that are likely to survive thymic selection. This idea was first proposed in \cite{Elhanati2014} based on the analysis of peripheral repertoires. Our analysis shows how the process unfolds and amplifies from initial receptor recombination ($P_{\rm gen}$) to the DP stage, and then further to each SP stage. 

Network analyses of repertoires have been successfully applied in a variety of contexts \cite{BenHamo2011,BashfordRogers2013,Madi2017,Pogorelyy2018b, ALICETCRNET,Miho2019,Ronel2021}. 
Following \cite{TCRNET}, we showed that diversity reduction is accompanied with the concentration of the network around the high-degree nodes. Remarkably, our models can accurately predict the changes in these local network structures. Our best-achieving model soNNia uses a nonlinear artificial neural network architecture, allowing it to capture complex interactions. However, we expect the high dimensionality of the sequence space to limit that potential, in particular, if the repertoire landscape is irregular or ``rugged,'' with deep valleys and hills. The usual view of thymic selection is that it should deplete specificities to self-antigens. Since antigen-specific TCRs form clusters in sequence space \cite{Dash2017,Glanville2017}, {\GI negative selection could have translated in the elimination of entire clusters corresponding to these forbidden self-antigen specific TCRs. This in turn should have created valleys and shaped a rugged landscape in the sequence space.}
The success of the model in the face of this potential issue is supported by our observation that the landscape is mostly ``smooth'', as demonstrated by the ability to approximate the probability of a sequence by that of its neighbours.
{\GI This highlights that the elimination of specific TCRs is a complex phenomenon, likely multifactorial, integrating not just specificities but also other parameters that control the level of T cell activation.} Understanding how these observations can be reconciled with the classical view of negative selection \cite{Camaglia2022} remains an interesting question for future work.

We further asked whether the local neighbourhood of a sequence carries information about its cell type. We defined a distance between repertoires based on the similarity of their neighborhood structure, which correlates well with the global, model-based Jensen-Shannon divergence, and recapitulates the hierarchy between the different stages of thymic development. While our neighbor-based distance is in principle model-free, in practice we could only evaluate it reliably using our data-driven generative models because of sampling issues. Nonetheless, our results suggest possible ways to use local network information to compare repertoire subsets, and to study their dependence on cell type, health condition, or age.

Local network structures can be used to detect responding clones during an infection, by looking for sequences with more neighbours than expected in a single repertoire \cite{ALICETCRNET}. Our results on comparing repertoires with a local neighborhood distance suggest that these local differences could also be used to identify sequences that are particularly enriched in one repertoire versus another. This could allow us to define sequences that are characteristic of particular repertoires, and use them to better understand the relationship between cell subset and TCR, with potential applications for diagnosis and phenotyping.

 \section{Methods}

\subsection{Code and data availability}
{\G{The code used to reproduce the figures can be found at \path{https://github.com/statbiophys/Global_and_Local_Variability_2023}}}

\subsection{Multidimensional Scaling}
The objective of Multidimensional scaling is to find a lower dimensional representation of data that preserves the similarity between samples in the dataset.  Given $D$ data points and precomputed distances $d_{ij}$ between points it infers data coordinates $(x_1,\dots,x_D)$ with $x_i\in \mathbb{R}^N$ and $N$ small by minimizing an objective function called stress:
\begin{equation}
Stress(x_1,\dots,x_D)=\sqrt{\sum_{i\neq j}(d_{ij}-||x_i-x_j||_2)^2}
\end{equation}

\subsection{Inference of a representation space for TCR sequences}
In order to better visualize the differences between repertoire subsets from different stages of  maturation in the thymus, we develop a method to map the receptor sequences in a representation space that carries information about selection. We build a feed-forward neural network {\GI that} outputs the selection factor for each subset and has a hidden layer of dimension 2; see Fig S2A for a sketch of the network architecture.  We infer selection factors by maximizing the joint objective
\begin{equation}
\mathcal{L}(\theta)= \sum_{t\in \mathcal{T}} \mathbb{E}_t[-E^t_\theta] -\log \mathbb{E}_{\mathcal{G}}[e^{-E^t_\theta}]
\end{equation}
where $\mathcal G$ is the set of sequences generated by the $P_{\gen}$ model, $\theta$ the model parameters, $E_\theta$ the energy that the model assigns to the sequence, and, with abuse of notation, we identify with $\mathcal{T}$ the cell types of the thymic samples and the corresponding dataset. It is important to clarify that this architecture is not an autoencoder because we are not trying to reconstruct the distribution of sequences, but we are only interested in characterizing the selection factor. For this reason, the representation space will carry information only about selection. We implement the model using the Keras software \cite{keras} and infer its parameters using the RMSprop stochastic gradient descent algorithm \cite{HINTON}.  After inference, we map all sequences to the two-dimensional hidden space.  As it can be seen in Fig.~S2B, we do not observe any clear separation between cell types in this representation space. On the other hand, the averages of the distributions are organized in a clear one-dimensional subspace that follows increasing selection.

\subsection{Smoothness of $P_{\rm post}$}
If we assume that the post-selection probability $P_{\rm post} = P_\gen Q$ does not vary significantly in a local region of sequence space, we can ask whether our estimators can predict the number of observed neighbours $N^{obs}(x)$ of a receptor sequence defined by higher cut-offs in {\GI the }hamming distance. In Fig.~S4A we show that the estimators $P^{\rm lin}_{\rm post}$ (linear selection model), $P^{\rm deep}_{\rm post}$  (deep selection model) and $P_{\rm gen}$ (pre-selection generation model) perform reasonably well also in this regime.  Surprisingly, there is no loss in performance for the threshold at hamming distance 2. For higher cut-offs the average performance decreases and the standard deviation increases. The estimators remain however significantly correlated with $N^{obs}(x)$.

The previous result is consistent with the idea that $P_{\rm post}$ is smooth in the neighbourhood of a given sequence. We are then motivated to push this assumption even further to define alternative estimators for $N^{obs}(x)$. The first approximation we can perform is to assume that all neighbours have similar probability, as it is done in the main text. This estimator is computationally more efficient as it requires only a single evaluation of the probability.

Alternatively, we can assume that  selection factors do not vary considerably within a neighbourhood and  approximate
$P_{\rm post}(nn_x) \simeq P_{\rm gen}(nn_x) Q(x)$ for the  post-selection models.  Since $P_{\rm gen}(nn_x)$  can be efficiently estimated via dynamic programming \cite{Sethna2019}, the resulting estimator turns out to be more efficient than the exact one, yet not as much as $P_{\rm post}(x)$. We compare the two approximations in Fig.~S4B for $P_{\rm post}^{\rm deep}$. Their performance is approximately comparable to the $P_{\rm post}^{\rm lin}(nn_x)$ for the smallest cut-off value. They generically perform better than  $P_{\rm gen}(nn_x)$ for all cut-off values that we tested. As expected the estimator that assumes smoothness only in selection $Q$ slightly outperforms the one that assumes smoothness of the whole probability $P_\post$.

In conclusion, we find that the three best choices for estimating the number of nearest neighbours are, $P^{\rm deep}_{\rm post}(x)$, $P_{\rm gen}(nn_x) Q^{\rm deep}_{\rm post}(x)$ and  $P_{\rm post}^{\rm deep}(nn_x)$. The three estimators have increasing performance but are also more computationally expensive: the choice of which one to use will thus depend on the specific application and the amount of available data.  Since these three estimators perform better at higher cut-offs in hamming distance than the original $P_{\rm gen}(nn_x)$ \cite{Pogorelyy2018b} at the smallest cut-off,  we expect that integrating information from higher cut-offs in hamming will increase the statistical power of enrichment analyses.

\subsection{Global Distance between repertoires $D_{\rm JS}$}
The Jensen-Shannon divergence $D_{\rm JS}$ is a symmetric measure between two probability distributions. It can be used to quantify the difference between any two repertoires defined by the post-selection probabilities $P_{\rm post}^r=\Q^{r}P_\gen^r$ and $P_{\rm post}^{r'}=\Q^{r'}P_\gen^{r'}$: 
\begin{equation}
D_{\rm JS}(r,r')=\frac{1}{2}\Big\langle \log_2 \frac{2\Q^{r}}{\Q^{r} +\Q^{r'}} \Big\rangle_{r}
+ \frac{1}{2} \Big\langle \log_2 \frac{2\Q^{r'}}{\Q^{r} +\Q^{r'}} \Big\rangle_{r'}
\label{eq:djs}
\end{equation}
where $\langle \cdot \rangle_{r}$ denotes the expectation value with respect to $P_{\rm post}^r$. In practice, we estimate $\langle \cdot \rangle_{r}$ as an empirical average by sampling 20,000 sequences from $P_{\rm post}^r$.

\subsection{Local Distance between repertoires $D_{\rm L}$}
The local distance $D_{\rm L}$ between repertoires, as defined in Eq.~\ref{eq:local_distance}, can be directly evaluated on  sequences shared between repertoires by computing the correlation between the observed number of neighbors in the respective datasets. In Fig.~S5A, we evaluate the $D_{\rm L}$  between independent subsamples from the same dataset. We show that for typical repertoire sizes that are found in the literature ($10^4$-$10^5$ unique receptor sequences), the estimator is too noisy to resolve the observed differences between repertoires (Figs.~3B,C), which can be as small as $D_L \sim 0.1~{\rm bits}$.  

In order to overcome this difficulty we evaluate the estimator using the expected number of neighbors $n_{\rm pred}$ computed by the models. We find that $D_{\rm L}$  converges quickly as a function of the number of sequences evaluated; see Fig.~S5B.  In the main analysis we evaluate $D_{\rm L}$ with 200 sequences sampled from each dataset.

As sequencing technologies improve, we expect the model-free estimator to successfully be used to compare the local structure of different repertoires.
As a proof of concept, we compare the model-free version of the $D_{\rm L}$ estimator to the global distance $D_{\rm JS}$ for the biggest datasets present in our data (6 datasets with at least 80,000 unique amino acid sequences) and we are able to reproduce the results of Fig.~3A,  see Fig.~S5C.

\section{Acknowledgement}
The study was supported by the European Research Council COG 724208 (AMW,  TM, GI), the ANR-19-CE45-0018 ``RESP-REP" (AMW,  TM) from the Agence Nationale de la Recherche, the DFG grant CRC 1310 ``Predictability in Evolution" (AMW,  AN, TM, GI), the National Institutes of Health MIRA award (R35 GM142795 to A.N.), the CAREER award from the National Science Foundation (grant No: 2045054 to A.N.), the MPRG funding through the Max Planck Society (AN,  GI), the ERC-Advanced TRiPoD (322856 to D.K.), LabEx Transimmunom (ANR-11-IDEX-0004-02 to D.K.), RHU iMAP (ANR-16-RHUS-0001 to D.K.), the European Research Area Network-cardiovascular diseases (ANR-18-ECVD-0001 to E.M.F.), and iReceptorPlus (H2020 Research and Innovation Programme 825821 to D.K. and E.M.F). This work benefited from equipment and services from the iGenSeq core facility at Institut du Cerveau (ICM, Paris, France) for all the data production. 

%

\renewcommand{\theequation}{S\arabic{equation}}
\renewcommand{\thefigure}{S\arabic{figure}}
\setcounter{figure}{0}

\appendix

\onecolumngrid

\newpage  
 \section{Supplementary Material}

\begin{figure*}[ht!]
\begin{center}
\includegraphics[width=1.\linewidth]{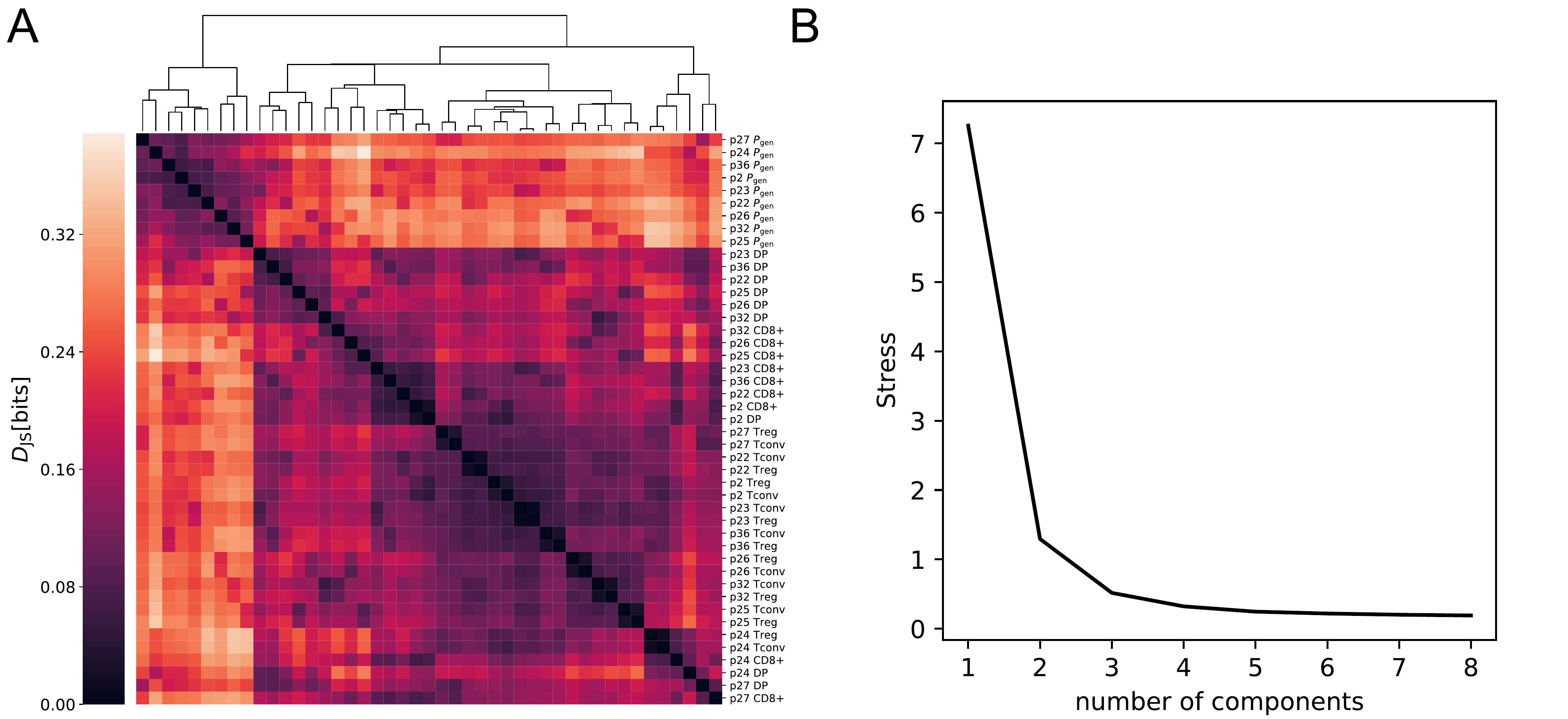}
\caption{ {\bf (A)} distance matrix $\mathcal{M}_g$ between different
  datasets.  Ordering is performed using the hierarchical clustering
  algorithm implemented in the seaborn clustermap function. {\bf (B)} Stress as a function of the number of
  components of the Multi-Dimensional Scaling embedding of the matrix
  (see Methods). }
\label{fig:sklatz1}
\end{center}
\end{figure*}
\begin{figure*}[ht!]
\begin{center}
\includegraphics[width=1.\linewidth]{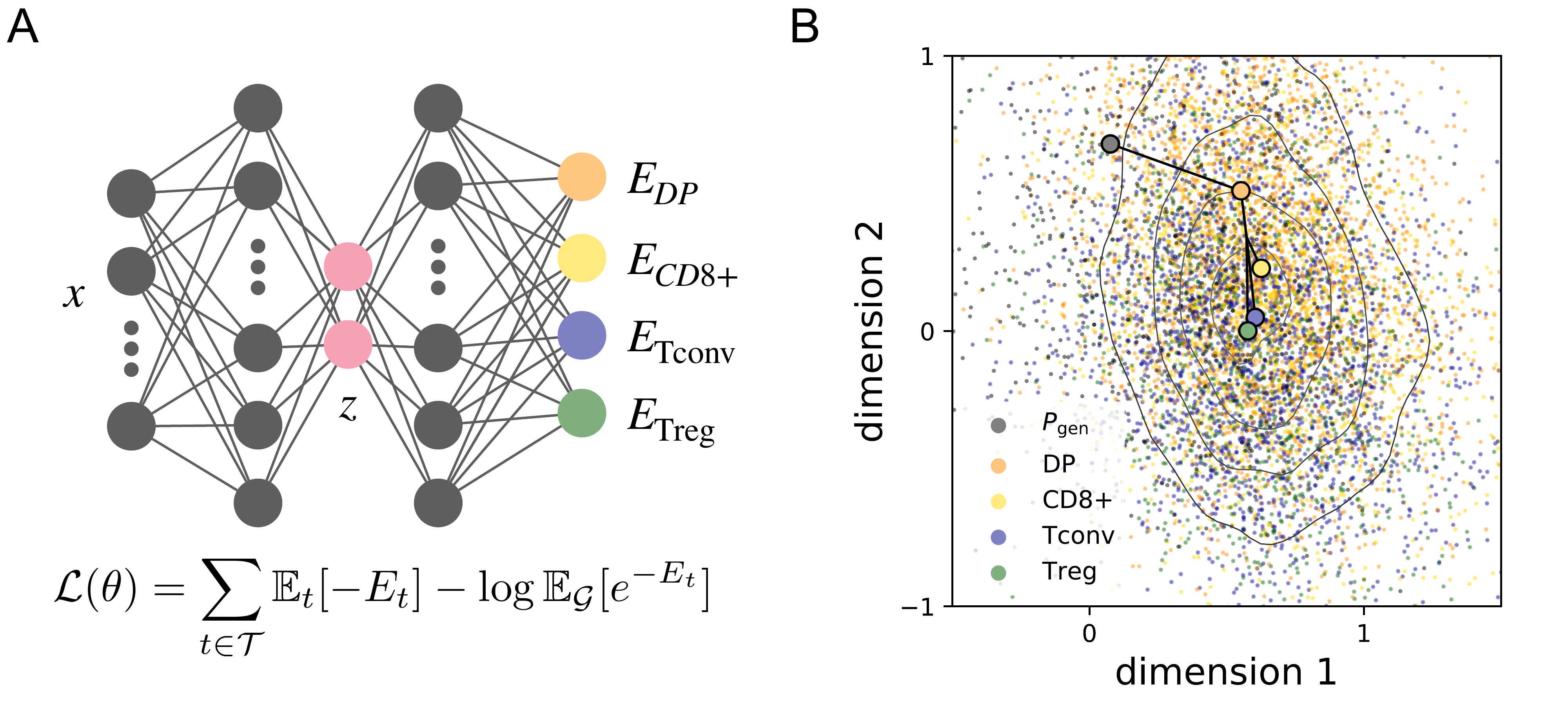}
\caption{{\bf (A)} Schematic shows the network architecture for joint inference of selection factors on a shared representation space of the same neural network. 
 {\bf (B)} Vizualization of the two-dimensional representation space $z$ where each point corresponds to a single sequence.
 No clear separation of subsets is visible in the representation space but the mean value of the distribution follows the differentiation process. 
 }
\label{fig:sklatz2}
\end{center}
\end{figure*}

\begin{figure*}[ht!]
\begin{center}
\includegraphics[width=1.\linewidth]{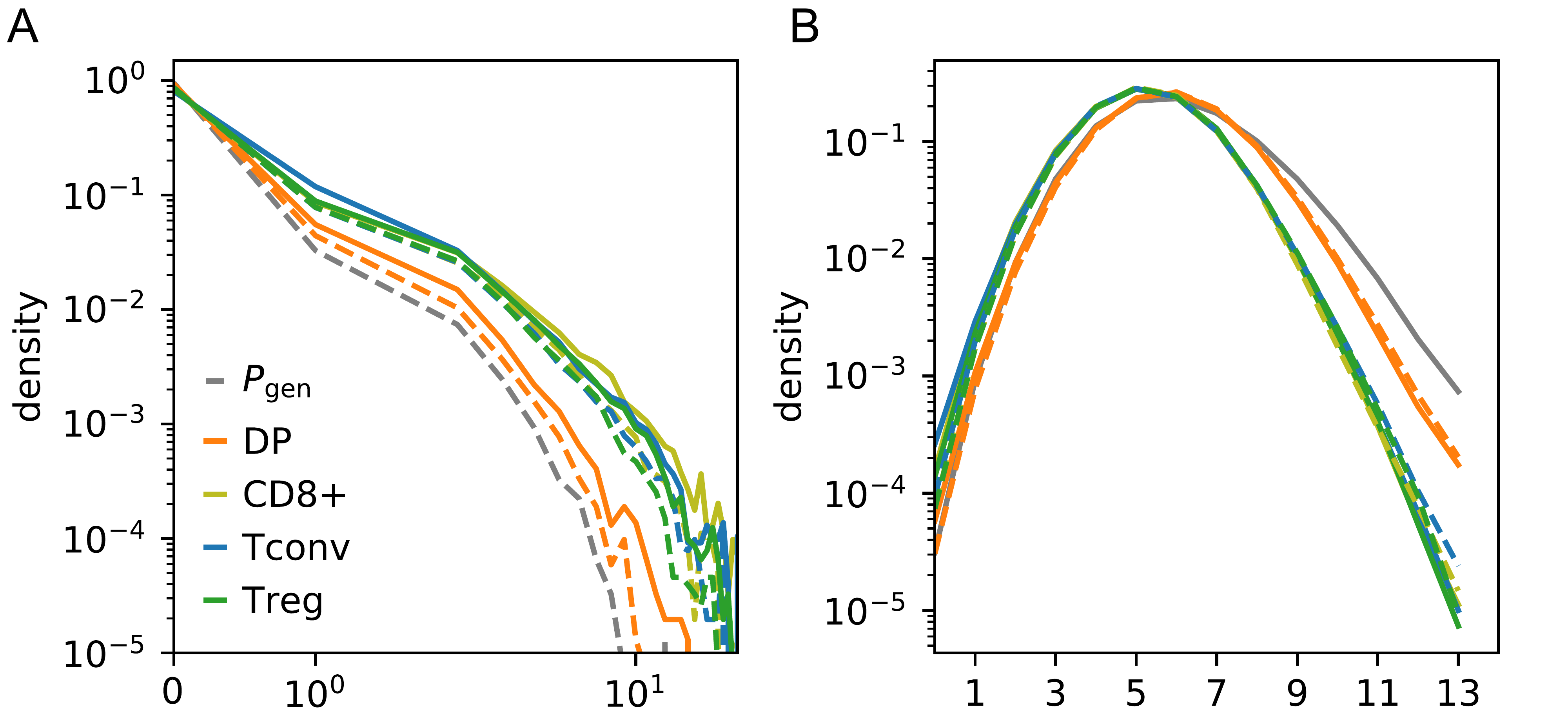}
\caption{{\bf Replication of main text Fig 2 with synthetic data.} {\bf(A)} The probability that a sequence has exactly $n$ nearest neighbours with the same $VJ$ genes is shown.  {\bf(B)} The probability that two randomly sampled sequences with the same $VJ$ gene combination have Hamming distance $d_H$ between their aminoacid sequences is shown.}
\label{fig:sklatz3}
\end{center}
\end{figure*}

\begin{figure*}[ht!]
\begin{center}
\includegraphics[width=1\linewidth]{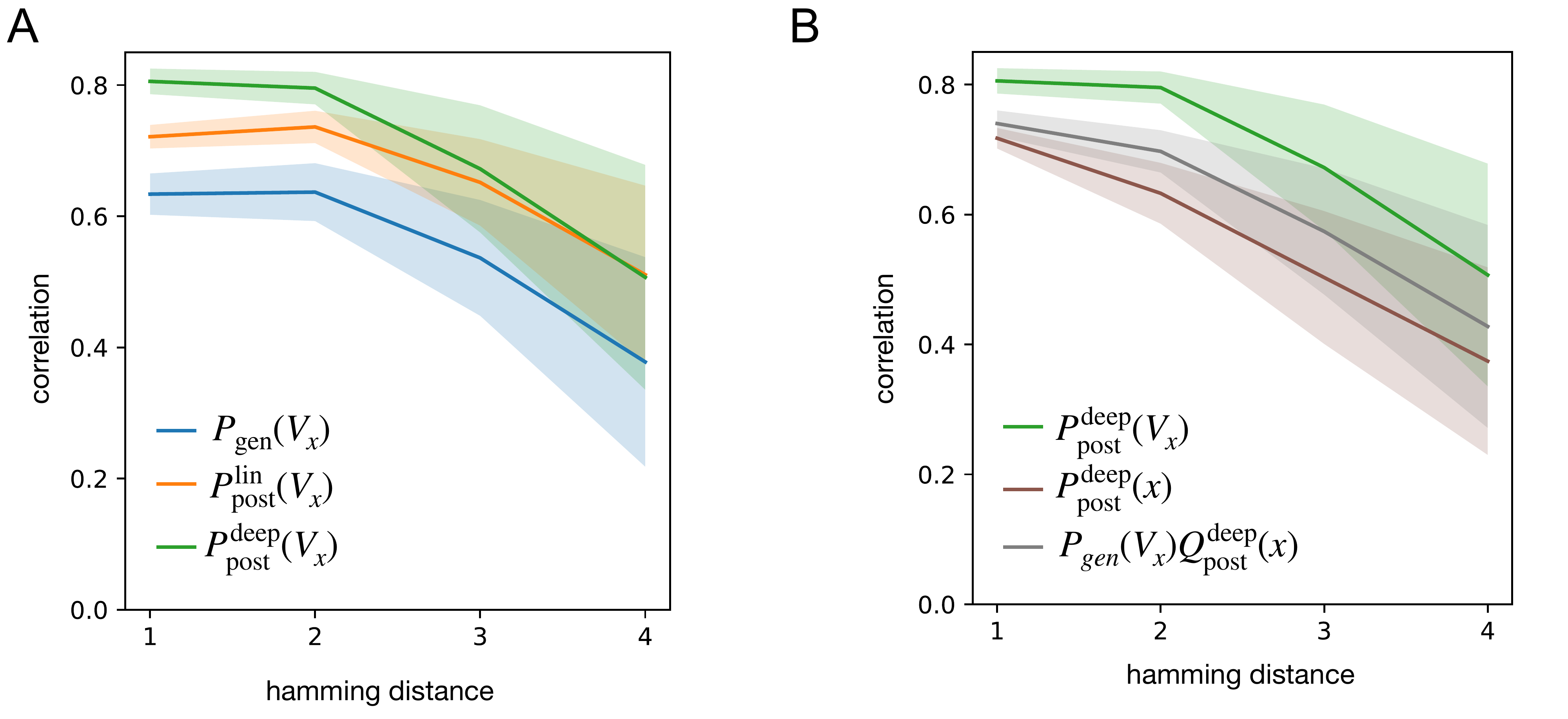}
\caption{{\bf Smoothness of $P_{\rm post}$.} {\bf(A)} Comparison of alternative estimators for the number of nearest neighbours defined by different cut-offs in the hamming distance. {\bf(B)}  We compare performance for different approximations of  $P_{\rm post}^{\rm deep}(V_x)$ which assume smoothness of the whole distribution, $P_{\rm post}^{\rm deep}(x)$,  or only of the selection factors,  $P_{\rm gen}(V_x) Q^{\rm deep}_{\rm post}(x)$. }
\label{fig:sklat4}
\end{center}
\end{figure*}

\begin{figure*}[ht!]
\begin{center}
\includegraphics[width=1\linewidth]{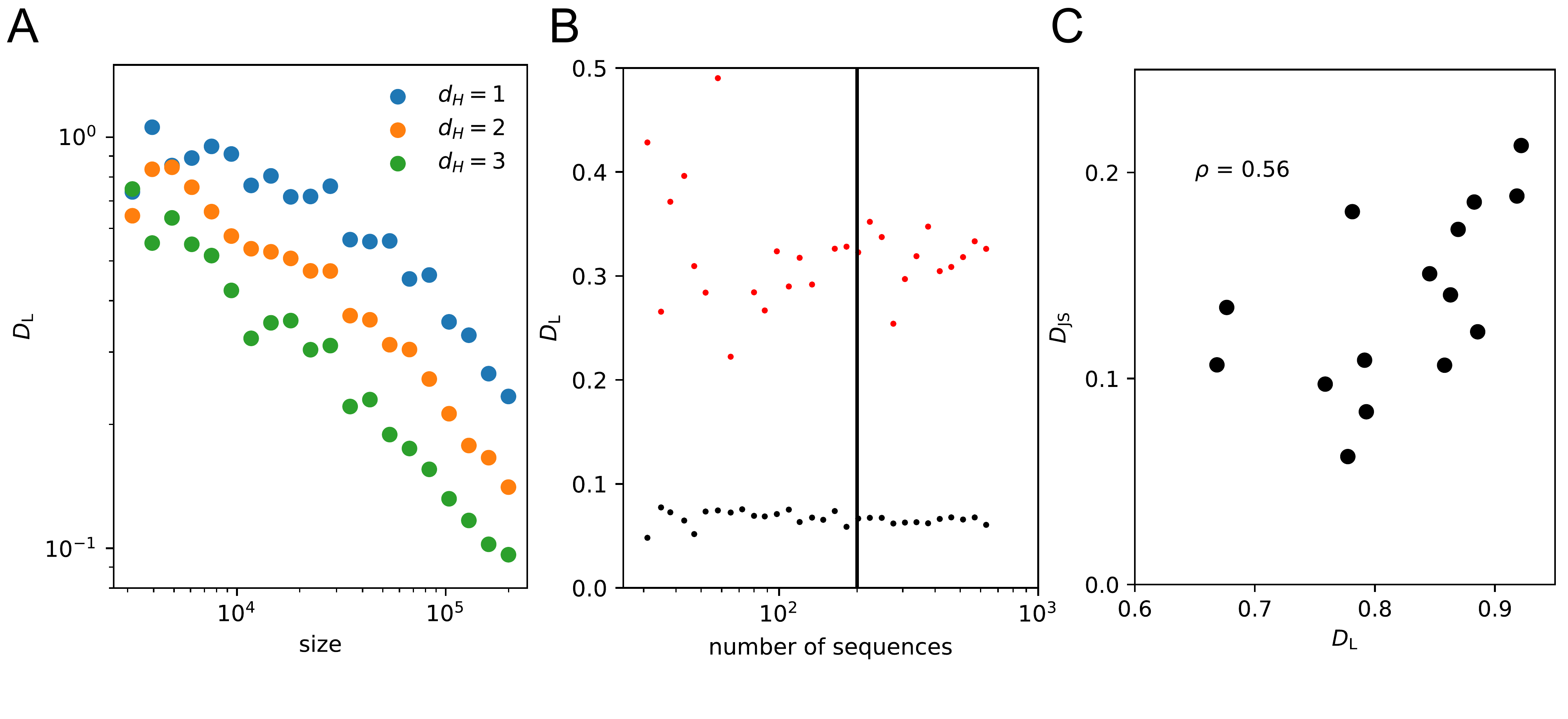}
\caption{ {\bf Local distance $D_{\rm L}$ between repertoires.} {\bf(A)} We estimate the noise in the model-free version of the estimator $D_{\rm L}$ by comparing subsamples from the same dataset at increasing sizes. We repeat the experiment for different cutoffs in Hamming distance ($d_H \in \{1,2,3\}$) that define the neighbourhood of a sequence. {\bf(B)} Evaluation of the probability for all neighbors of a sequence is computationally intensive.  We evaluate the convergence of the model-based version of the estimator $D_{\rm L}$ as a function of how many sequences are evaluated. We observe that for comparisons between similar repertoires (black dots) and very different repertoires (red dots), the estimator converges around $10^2$ evaluated sequences.  We highlight with a vertical line the sample size (200 sequences) used to produce the results in the main text. {\bf(C)} Replication of the results in Fig.~3A with the model-free version of the estimator for the largest datasets present in our study (six datasets with at least 80,000 unique receptor sequences). }
\label{fig:sklat5}
\end{center}
\end{figure*}

\begin{figure*}[ht!]
\begin{center}
\includegraphics[width=1\linewidth]{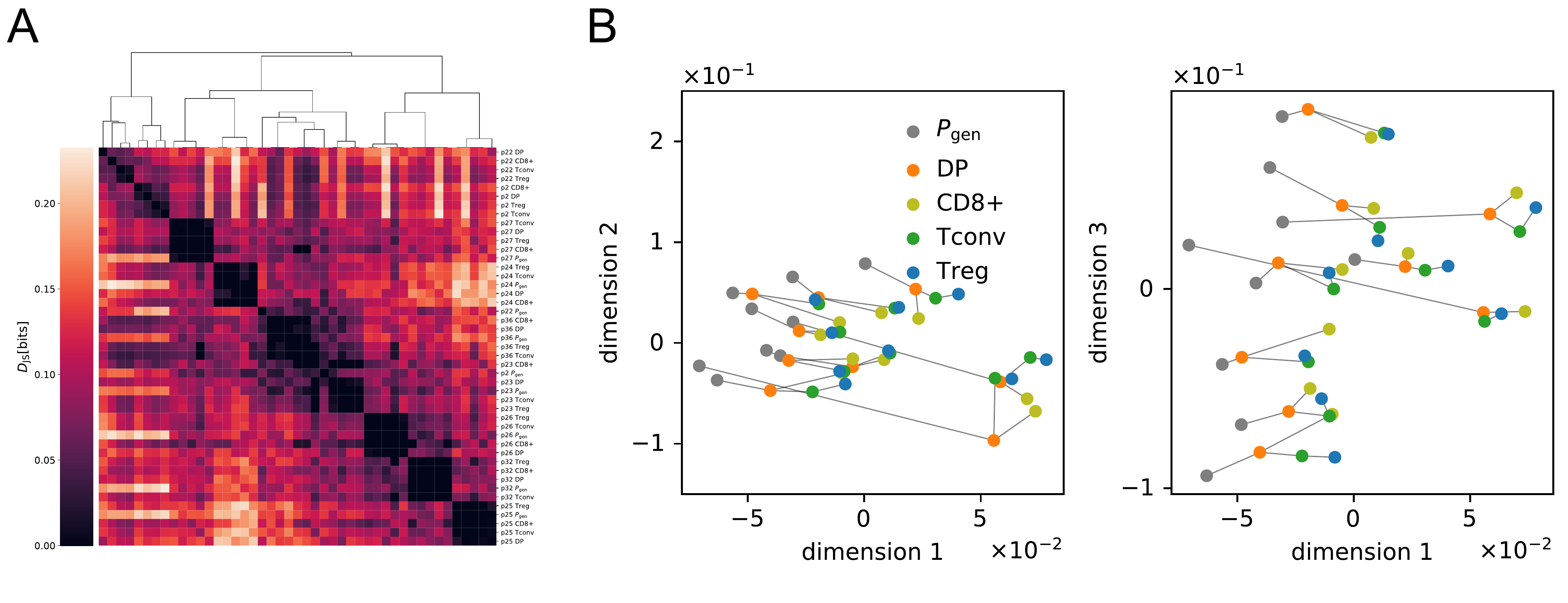}
\caption{\G{ \bf Aminoacid Composition Matters.} Replication of Fig.~S1A {\bf(A)} and Fig.~1D {\bf(B)} by inferring selection models without aminoacid features. As clearly shown in the distance matrix, the VJ gene features cannot accurately separate intra-individual cell subsets.}
\label{fig:sklat6}
\end{center}
\end{figure*}

\begin{figure*}[ht!]
\begin{center}
\includegraphics[width=0.5\linewidth]{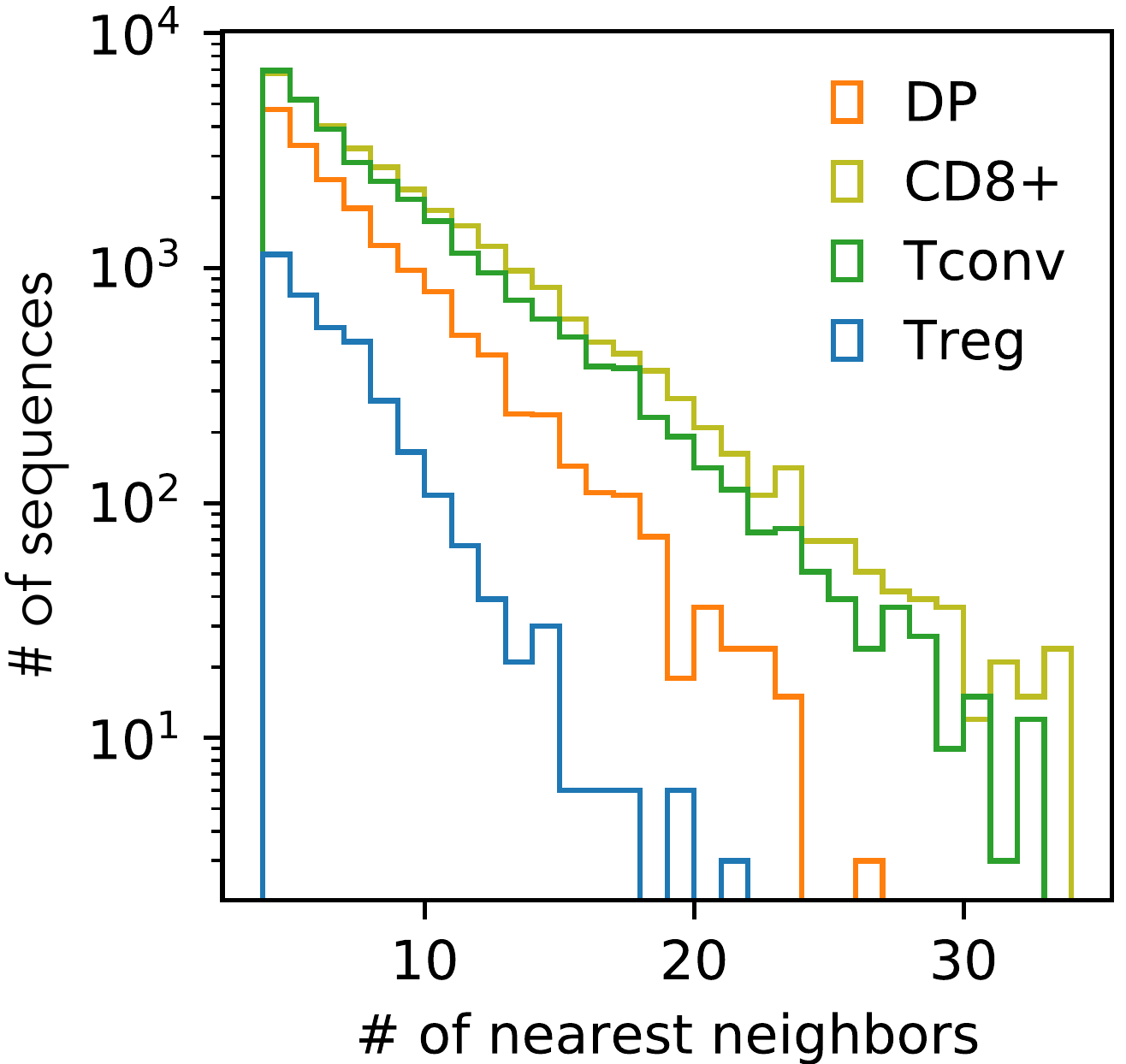}
\caption{ Histogram of the number of nearest neighbours for each cell type. Due to the smaller dataset size, TCRs in the Treg and DP subset have fewer nearest neighbours on average, resulting in a nosier comparison to the model in Fig.~2E-F.}
\label{fig:sklat7}
\end{center}
\end{figure*}

\end{document}